\documentclass[prb,showpacs,loatfix,twocolumn,superscriptaddress,byrevtex]{revtex4}
\usepackage{amsmath}
\usepackage{graphicx}
\usepackage{float}
\usepackage{bm}
\begin{document}

\bibliographystyle{apsrev}

\title{Multiple bosonic mode coupling in the charge dynamics of the
electron-doped superconductor \boldmath (Pr$_{2-x}$Ce$_x$)CuO$_4$
\unboldmath}
\author{E. Schachinger}
\email{schachinger@itp.tu-graz.ac.at}
\affiliation{Institute of Theoretical and Computational Physics,
Graz University of Technology, A-8010 Graz, Austria}
\author{C. C. Homes}
\affiliation{Condensed Matter Physics and Materials Science Department,
Brookhaven National Laboratory, Upton, NY 11973, USA}
\author{R. P. S. M. Lobo}
\affiliation{Laboratoire Photons et Mati\`ere (CNRS UPR 5), LPS-ESPCI,
Universit\'e Pierre et Marie Curie, 10 rue Vauquelin, F-75231 Paris Cedex 5,
France}
\author{J. P. Carbotte}
\affiliation{Department of Physics and Astronomy, McMaster University, Hamitlon,
Ontario N1G 2W1, Canada}
\affiliation{The Canadian Institute for Advanced Research, Toronto, Ontario M5G
1Z8, Canada}
\date{\today}
\begin{abstract}
We analyze optical spectroscopy data of the electron-doped superconductor
(Pr$_{2-x}$Ce$_x$)CuO$_4$ (PCCO) to investigate the coupling of the charge
carriers to bosonic modes. The method of analysis is the inversion of the
optical scattering rate $\tau^{-1}_{\rm op}(\omega,T)$ at different
temperatures $T$ by means of maximum entropy technique combined with Eliashberg
theory. We find that in the superconducting state the charge carriers couple to
two dominant modes one at $\sim 12\,$meV and a second one at $\sim 45\,$meV
as well as to a high energy background.
The low energy mode shows a strong temperature dependence and disappears at or
slightly above the critical temperature $T_c$. The high energy mode exists
above $T_c$ and moves towards higher energies with increasing temperatures. It
becomes less prominent at temperatures $> 100\,$K above which it evolves
into a typical spin-fluctuation background. In contrast to the hole-doped
High-$T_c$ superconductors PCCO proves to be a superconductor close to the
dirty limit.
\end{abstract}
\maketitle
%

%
%
\section{Introduction}
The optical scattering rate $\tau^{-1}_{\rm op}(\omega)$ in the infrared regime
evolved into an important tool to extract microscopic information on the
coupling of charge carriers to bosonic modes in the form of the
electron-exchange boson interaction spectral density $I^2\chi(\omega)$ in the
High-$T_c$ cuprates. The definition of the optical scattering rate itself is
based on a generalized Drude form valid for correlated electron systems:
\begin{equation}
  \label{eq:1}
  \sigma(\omega,T) = \frac{i\Omega^2_p}{4\pi}\frac{1}
  {\omega-2\Sigma_{\rm op}(\omega,T)}.
\end{equation}
This equation relates the complex optical self energy $\Sigma_{\rm
op}(\omega,T)$ at a given temperature $T$ to the complex optical conductivity
$\sigma(\omega,T)$. The optical scattering rate is related to the imaginary
part of the optical self energy $\Sigma_{2,\rm op}(\omega,T)$ via
\begin{equation}
  \label{eq:2}
  \tau^{-1}_{\rm op}(\omega,T) = -2\Sigma_{2,\rm op}(\omega,T)
  = \frac{\Omega_p^2}{4\pi}\Re{\rm e}\left\{\sigma^{-1}(\omega,T)
  \right\}.
\end{equation}
Here $\Omega_p$ is the plasma frequency and $\omega$ an energy
variable.

It was first demonstrated by Marsiglio {\it et al.}\cite{mars1} that there
exists an approximate relation between $I^2\chi(\omega)$ and $\tau^{-1}_{\rm
op}(\omega)$ of the form:
\begin{equation}
  \label{eq:3}
  I^2\chi(\omega)\simeq W(\omega) = \frac{1}{2\pi}
  \frac{d^2}{d\omega^2}\left[\frac{\omega}{\tau_{\rm op}(\omega)}
  \right].
\end{equation}
This relation is valid only at low temperatures and up to energies at which
$W(\omega)$ first becomes negative, i.e. unphysical. This second derivative
method was used by Carbotte {\it et al.}\cite{carb3} to demonstrate that in the
optimally doped system YBa$_2$Cu$_3$O$_{6+\delta}$ (YBCO) the quasiparticles
couple to a boson resonance at $41\,$meV which corresponds to a spin one
resonance observed by neutron scattering in the imaginary part of the spin
susceptibility.\cite{ross,dai} Moreover, the temperature dependence of this
resonance peak was found to be identical to the $T$ variation of the equivalent
structure in the spectral function $I^2\chi(\omega)$ derived from
optics.\cite{schach1} Finally, Schachinger and Carbotte\cite{schach2} predicted
such a spin one resonance to exist in the thallium compound
Tl$_2$Ba$_2$CuO$_{6+\delta}$ (Tl2201). This was later confirmed by He {\it et
al.}\cite{keimer} using neutron scattering.

The next step in the development of methods which help to extract information
on the electron-exchange boson interaction spectral density $I^2\chi(\omega)$
from optics, was made by Dordevic {\it et al.}\cite{dorde} who developed a new
method based on inverse theory. These authors concentrated on the approximate
relation
\begin{equation}
  \label{eq:4}
  \tau^{-1}_{\rm op}(\omega,T) = \tau^{-1}_{\rm imp}+
  \int\limits_0^\infty\,d\nu\,K(\omega,\nu;T)
  I^2\chi(\nu)
\end{equation}
reported by Shulga {\it et al.}\cite{shulga1} Eq.~\eqref{eq:4} is based on
Eliashberg theory and is valid in the normal state. Here $\tau^{-1}_{\rm imp}$
is an energy independent impurity scattering rate and the kernel
$K(\omega,\nu;T)$ is given by:
\begin{eqnarray}
  \label{eq:5}
  K(\omega,\nu;T)  &=&  \frac{\pi}{\omega} \left[ 2\omega\text{coth}
  \left(\frac{\nu}{2T}\right) - (\omega+\nu)\text{coth}\left(
  \frac{\omega+\nu}{2T}\right)\right.\nonumber\\
  &&\left.+(\omega-\nu) \text{coth} \left(\frac{\omega-\nu}{2T}\right) \right].
\end{eqnarray}
Relation \eqref{eq:4} was extended by Carbotte and Schachinger\cite{carb2} to
the superconducting state of a $d$-wave superconductor at $T=0$ using the
kernel
\begin{eqnarray}
  \label{eq:6}
  K(\omega,\nu;T=0) &=& \frac{2\pi}{\omega}\bigg\langle(\omega-\nu)
  \theta[\omega-2\Delta(\vartheta)-\nu]\nonumber\\
  &&\times\left.
     E\left(\sqrt{1-\frac{4\Delta^2(\vartheta)}{(\omega-\nu)^2}}\right)
  \right\rangle_\vartheta.
\end{eqnarray}
This kernel is based on a clean limit, i.e.: $\tau^{-1}_{\rm imp} =0$,
perturbation theory expansion of BCS theory reported by P.B. Allen\cite{allen}
for an $s$-wave superconductor which has also been considered by Dordevic {\it
et al.}\cite{dorde} In as much as one can think of a $d$-wave superconductor as
a superposition of $s$-wave with variable gaps, Eq.~\eqref{eq:6} follows as a
first approximate generalization of Allen's work to $d$-wave. Here
$\langle\cdots\rangle_\vartheta$ denotes the $\vartheta$-average which can be
limited to the interval $\vartheta\in[0,\pi/4]$ for symmetry reasons.
Furthermore, $\Delta(\vartheta) = \Delta\cos(2\vartheta)$ reflecting the
$d$-wave symmetry of the superconducting order parameter. Eq.~\eqref{eq:6}
ensures that the optical scattering rate is finite in the superconducting state
for $\omega>0$. Finally, $E(x)$ is the  complete elliptic integral of second
kind and $\theta(x)$ is the step function.

Eq.~\eqref{eq:6} explains why in the superconducting state of a $d$-wave
superconductor the signatures of the $I^2\chi(\nu)$ spectral density in the
optical scattering rate $\tau_{\rm op}^{-1}(\omega)$ are displaced at least by
the gap amplitude $\Delta$.\cite{aban,schach10,tu} This is in contrast to a
classical $s$-wave superconductor in which those signatures are displaced by
$2\Delta_0$, with $\Delta_0$ the gap-edge. Finally, we would like to point out
that according to Eq.~\eqref{eq:4} non-zero contributions to the positive
definite spectral density $I^2\chi(\nu)$, i.e.: non-zero coupling of charge
carriers to an exchange boson at energy $\nu$, will always result in an
increase of the optical scattering rate because both kernels, Eqs.~\eqref{eq:5}
and \eqref{eq:6}, are positive definite. Thus, depressions or pronounced peaks
in the optical scattering rate $\tau_{\rm op}^{-1}(\omega)$ cannot be explained
by electron-exchange boson interaction and indicate additional influences like,
for instance, infrared activated phonons or quasiparticle density of states
effects.

The spectra derived from the deconvolution of Eq.~\eqref{eq:4} still contained
negative and, thus, unphysical parts. Nevertheless, it was demonstrated by
Dordevic {\it et al.}\cite{dorde} that the application of a deterministic
constraint can result in the regularization of the problem which allows to
the removal of unphysical negative values from the inverted spectral function
$I^2\chi(\nu)$. The authors also demonstrated that such a regularization can
result in a reduced quality of data reproduction.

In a final step, so far, in the development of inversion techniques it was
demonstrated by Schachinger {\it et al.}\cite{schach3} that the application of
maximum entropy techniques\cite{jaynes} maximizing the generalized
Shannon-Jaynes entropy\cite{sivia}
\begin{equation}
  \label{eq:6a}
  S = \sum\limits_{j=1}^N\left[a_j-m_j-a_j\log\frac{a_j}{m_j}
  \right]
\end{equation}
to deconvolute Eq.~\eqref{eq:4} will result in a positive definite spectral
density $I^2\chi(\nu)$ ensuring best possible data reproduction. Here, the
entropy $S$ measures the distance of a candidate vector ${\bf a} = \{a_j\vert
j=1,\ldots,N\}$ from the default model vector ${\bf m}=\{m_j\vert
j=1,\ldots,N\}$ which represents the most probable solution prior to the
observation of any data. In case of insufficient background information it
should be chosen constant, i.e.: $m_j = const,\, \forall j$, as will be done
throughout this paper.

To compensate for the approximate nature of Eq.~\eqref{eq:4} an additional
least squares fit procedure based on Eliashberg equations which have been
extended to $d$-wave superconductors (see Appendix~\ref{app:A}) proved,
finally, to be very successful in the inversion of optical
Bi$_2$Sr$_2$CaCu$_2$O$_{8+\delta}$ (Bi2212) data at various temperatures and
doping levels.\cite{hwang}

All this research resulted in one common denominator: in YBCO, Bi2212, and
Tl2201 the optical data suggest coupling of the charge carriers to a pronounced
boson mode, an `optical' resonance, at energies which agree in most cases with
the energies at which a spin one resonance is found by neutron scattering in
the imaginary part of the spin susceptibility. In a new experiment Vignolle
{\it et al.}\cite{vigno} showed, using neutron scattering, that the imaginary
part of the spin susceptibility in optimally doped (La$_{2-x}$Sr$_x$)CuO$_4$
(LSCO) develops two peak structures, a resonance at the low energy of $12\,$meV
while the second peak was found at $\sim 50\,$meV.
A maximum entropy analysis of
optical data reported by Gao {\it et al.}\cite{gao} on epitaxial optimally
doped LSCO thin films confirmed that, indeed, the electron-boson spectral
density $I^2\chi(\omega)$ develops two peaks at the energies reported by
neutron scattering.\cite{hwang1} This is also in agreement with earlier
results suggested by Zhou {\it et al.}\cite{zhou} from angular resolved photo
emission spectroscopy (ARPES) experiments in underdoped samples. They
interpreted their results as a coupling of the charge carriers to phonons
which is in some contrast to the results reported by
Vignolle {\it et al.}\cite{vigno} and Gao {\it et al.}\cite{gao}

Neutron scattering experiments by Wilson {\it et al.}\cite{wilson} report
the existence of a spin one resonance in the imaginary part of the
spin susceptibility of (Pr$_{1-x}$LaCe$_x$)CuO$_{4-\delta}$ (PLCCO)
for $x=0.12$ with its peak at $\sim 11\,$meV. It is centered around
$(\pi/2,\pi/2)$ in the two dimensional CuO Brillouin zone as is the
case in all other High-$T_c$ superconductors and it disappears at
temperatures above $T_c$. The doping $(x)$ dependence of this peak
was studied by M. Fujita {\it et al.}\cite{Fujita2008}
Scanning tunneling microscope (STM)
experiments performed on the same material by Niestemski {\it et al.}%
\cite{niest} support the existence of such a peak at $\sim 10\,$meV. As optical
data are available in the similar system (Pr$_{2-x}$Ce$_x$)CuO$_{4-\delta}$
(PCCO) we concentrate in this paper on the task to extract information on the
electron-boson spectral density $I^2\chi(\omega)$ from optical scattering rates
measured by Homes {\it et al.}\cite{homes} on a PCCO single crystal with
$x=0.15$ ($T_c = 20\,$K) and from similar data reported by Zimmers {\it et
al.}\cite{lobo1,lobo2} on thin, epitaxially grown PCCO films with $x=0.15$
($T_c = 21\,$K).

In Sec.~\ref{sec:2} we discuss the samples used for the analysis
in their main features. Data extraction techniques will also be
discussed shortly within this section. Section~\ref{sec:3}
concentrates on the analysis of the optical data and the discussion
of the important features of the $I^2\chi(\omega)$ spectra derived
from optical data. Finally, Sec.~\ref{sec:4} gives a short summary.
An appendix has been included to provide the necessary background
information on the extended $d$-wave Eliashberg formalism used for
impure systems.

%
%
%
%
\section{Materials}
\label{sec:2}
\subsection{The optimally doped \boldmath ($x=0.15$) \unboldmath PCCO single crystal}
\label{sec:2a}
We concentrate first on the experimental results reported by Homes {\it et
al.}\cite{homes} for an optimally doped PCCO single crystal ($x=0.15$, $T_c
\sim 20\,$K). The authors report two plasma frequencies, the first value
$\Omega_p = 13\,000\pm200\,$cm$^{-1}$ ($1.64\,$eV) reflects only those carriers
which participate in coherent ``Drude'' transport, and a larger value for
$\Omega_p = 19\,300\,$cm$^{-1}$ ($2.4\,$eV), which was determined from a
modification of the finite-energy $f$-sum rule.\cite{smith,hwang04} The two
values of $\Omega_p$ obtained using these different approaches indicate that
$\Omega_p$ is not well known in this material, and that its value depends
heavily on the method chosen to extract it from the experiment. This leads to
some uncertainty because according to Eq.~\eqref{eq:2} the plasma frequency
sets the scale of the optical scattering rate.
%
%
\begin{figure}[pt]
  \vspace*{-5mm}
  \includegraphics[width=9cm]{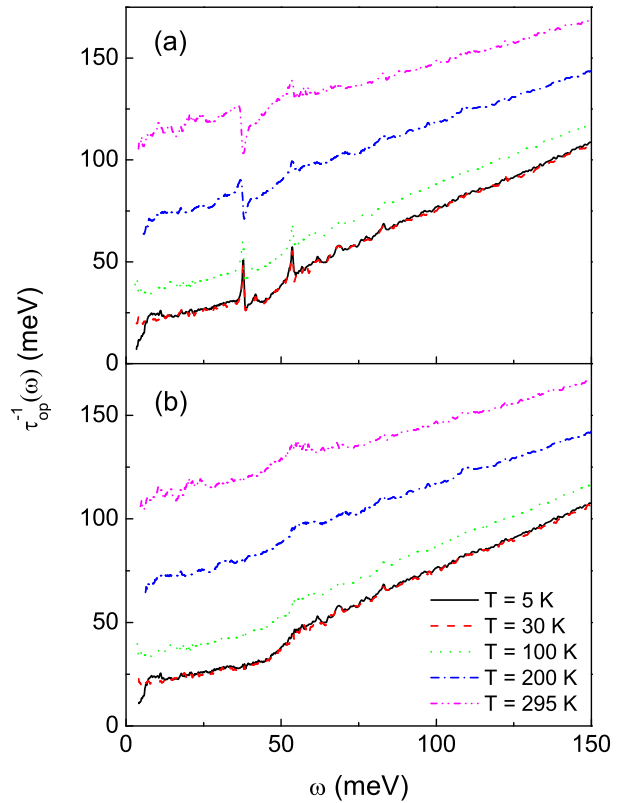}
  \vspace*{-1.0cm}
  \caption{(Color online) The optical scattering rate $\tau_{op}^{-1}(\omega)$
  in meV in an optimally-doped ($x=0.15$) thin PCCO single crystal ($T_c =
  20$~K) for several temperatures above and below $T_c$ with (a) the three
  infrared-active $E_u$ modes at $\approx 306$, 338, and 434~cm$^{-1}$
  ($\approx 37.9$, 41.9, and $53.8\,$meV) present,
  and (b) with the same phonons removed.}
  \label{fig:1}
\end{figure}
The smaller value of $\Omega_p$ finds its justification in a two-component
system where the Drude response is not directly connected to the mid-infrared
region which is conceived to be due to another band of electrons; this was the
approach used in the original analysis of the PCCO single crystal
data.\cite{homes}  However, in this work we use a one-component approach and as
a result the larger value of $\Omega_p$ is more appropriate. In this view Drude
and infrared regions come from the same electrons with the infrared part coming
from the incoherent boson-assisted processes, and the Drude originates from the
coherent quasiparticle response part of the carrier spectral function.
Following this argument, the experimental $\tau^{-1}_{\rm op}(\omega)$ data has
been derived from the raw reflectance data using $\Omega_p=2.4\,$eV together
with the dielectric constant at infinity, $\epsilon_\infty$, set equal to four.
It is worth remarking at this point that the dielectric function,
$\epsilon(\omega)$, and the
conductivity are related through the expression $\sigma(\omega) =
-i\omega[\epsilon(\omega) - \epsilon_\infty]/4\pi$.  Thus, $\epsilon_\infty$ is
required to derive the imaginary part of the optical conductivity, and as a
consequence the optical scattering rate is also influenced by this important
parameter.

The optical scattering rate at the lowest normal state temperature reported
($30\,$K) has a zero frequency offset of $\sim 22\,$meV ($\sim 177\,$cm$^{-1}$)
which indicates substantial contributions from impurity scattering.  In
addition, there are sharp features in the scattering rate shown in
Fig.~\ref{fig:1}(a) at approximately 306, 338, and 434~cm$^{-1}$
($\approx 37.9$, 41.9, and $53.8\,$meV) which have been
identified as the infrared-active $E_u$ lattice vibrations in this
material.\cite{homes00, homes02, braden05,homes} In order to remove this sharp
structure from the scattering rate, the dielectric function has been modeled
using a series of Lorentz oscillators
\begin{equation}
  \epsilon(\omega)_{osc}=\sum_j
  \frac {\omega_{p,j}^2} {\omega_j^2-\omega^2-i\gamma_j\omega},
  \label{eq:lorentz}
\end{equation}
where $\omega_j$, $\gamma_j$ and $\omega_{p,j}$ are the frequency, width and
effective plasma frequency of the $j$'th vibration, respectively.
The oscillator parameters
have been determined from fits to the real part of the optical conductivity
using a simple linear background; the fitted oscillator values are then used to
generate the phonon contribution to the complex dielectric function which is in
turn subtracted from the experimentally determined values for the real and
imaginary parts of the dielectric function, allowing the sharp phonon features
to be removed from the optical scattering rate, as shown in
Fig.~\ref{fig:1}(b).

%
%
\subsection{The optimally doped \boldmath ($x=0.15$) \unboldmath PCCO thin film}
\label{sec:2b}

We also analyzed data published on thin films by Zimmers
{\it et al.}\cite{lobo1,lobo2} and concentrate on the $x=0.15$ sample. 
The reflectivity of a thin film also contains information on the optical
properties of the underlying substrate. Hence one cannot obtain the film
optical functions using a straight-forward Kramers-Kronig inversion. In the
case of the PCCO films, the substrate response was determined separately.
The reflectivity of the total (film plus substrate) system was then fitted
by constructing a suitable Drude-Lorentz dielectric function for the
film.\cite{sant} One should note that the parameters used in this
Drude-Lorentz dielectric function are not to be taken as representative of
intrinsic excitations. They should be looked at as a convenient
parameterization of the material dielectric function. 
Once this dielectric function is known, any other optical function can be
generated straightforwardly. 

\begin{figure}[pt]
  \vspace*{-3mm}
  \includegraphics[width=9cm]{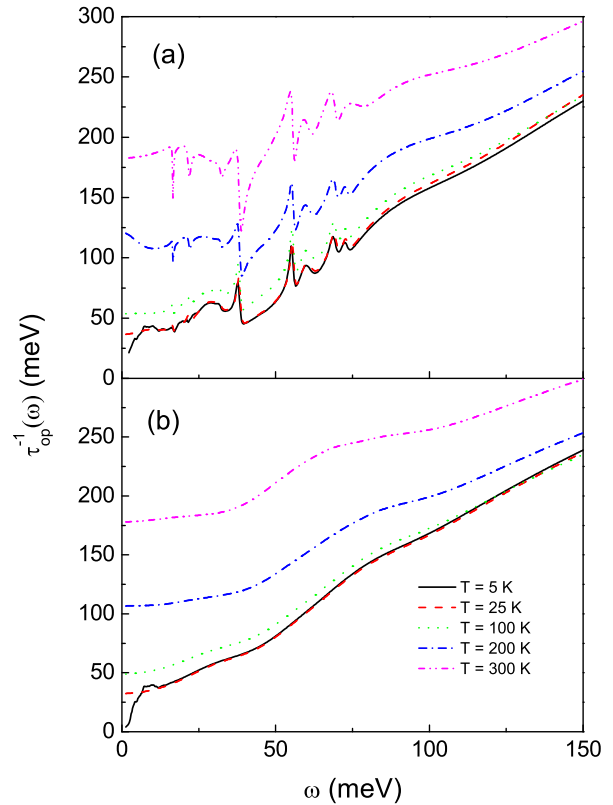}
  \caption{(Color online) The optical scattering rate
$\tau_{\rm op}^{-1}(\omega)$ im meV in an optimally doped ($x=0.15$)
PCCO thin film ($T_c=21\,$K) for several temperatures above and below
$T_c$ with (a) the infrared active phonon modes present and (b)
with these phonons removed.
}
  \label{fig:2}
\end{figure}

Figure~\ref{fig:2}(a) presents the optical scattering rate
$\tau^{-1}_{\rm op}(\omega)$ vs $\omega$ for the optimally doped, epitaxially
grown thin PCCO film ($x=0.15$, $T_c=21\,$K). The plasma frequency,
calculated from a finite sum rule, is
$\Omega_p = 17\,570\,$cm$^{-1}$ ($2.18\,$eV) and $\epsilon_\infty$ 
was set to four. Figure~\ref{fig:2}(b) shows the scattering rate once phonons 
have been eliminated.
%
%

The zero frequency offset of the optical scattering rate $\tau^{-1}_{\rm op}(0)
= 32.5\,$meV ($\sim 262\,$cm$^{-1}$) at $T=25\,$K which suggests a much higher
impurity concentration in comparison to the single crystal discussed in the
previous subsection. The overall shape of the scattering rate in the
$x= 0.15$ film is similar to the crystal response. The phonons in the film
tend to be broader and less screened by free carriers than in the crystal.
These effects are consistent with a larger disorder and smaller weight of the
coherent peak in the film, as we describe below.

There is a scaling by roughly a factor of two required if one wants to compare
directly the $x=0.15$ film and crystal data. It comes from
the difference in the absolute value of their respective scattering rates. It
is worth noticing that this is not an effect originating in the scattering
rate of carriers participating in the coherent transport. As discussed in
Sec.~\ref{sec:2a}, taking into account the coherent transport alone, the low
frequency optical conductivity of the $x=0.15$ single crystal at $T=30\,$K can
be described by a Drude peak having $\Omega_p = 1.61\,$eV
($13\,000\,$cm$^{-1}$) and $\tau^{-1}_{\rm op}(\omega=0) \approx 11\,$meV
($90\,$cm$^{-1}$). Performing the same analysis on the $x=0.15$ film at
$T=25\,$K one obtains $\Omega_p = 1.18\,$eV ($9\,500\,$cm$^{-1}$) and
$\tau^{-1}_{\rm op}(\omega=0) = 11\,$meV. The scattering rate is the same but
the weight of the Drude peak in the film is significantly smaller. As the film
and the crystal compositions are nominally the same, the \textit{f}-sum rule
states that the area under their respective optical conductivities should be
the same. Hence, the weight lost in the coherent peak of the film is
redistributed as an incoherent background. This effect will produce, as
observed, a higher mid-infrared optical conductivity in the film and hence a
broader incoherent scattering rate. Thus, a factor of two is not far from the
difference in dc-conductivity [$\sigma_1(\omega)$ extrapolated to $\omega=0$]
observed in these two samples ($3 \times 10^4\,\Omega^{-1}$\,cm$^{-1}$ for the
crystal; $1.7 \times 10^4\,\Omega^{-1}$\,cm$^{-1}$ for the film).

\section{Data analysis and discussion}
\label{sec:3}

\subsection{The \boldmath $x=0.15$ \unboldmath PCCO single crystal}
\label{sec:3a}

Because of the rather large normal state zero frequency offset of
the optical scattering rate PCCO can
no longer be treated in the clean limit as has been done so far for
the systems Bi2212, LSCO, Tl2201, and YBCO.
Thus, the impurity scattering rate
$\tau^{-1}_{\rm imp}$ gains importance in Eq.~\eqref{eq:4} and has to be
treated as an external parameter in the maximum entropy deconvolution
of this equation.

The complete inversion procedure consists of two steps. First Eq.~\eqref{eq:4}
together with the appropriate kernel is deconvoluted using a classical
maximum entropy method. The default model is set to a constant which
is adjusted from temperature to temperature to ensure best possible
data reproduction at high energies $\omega > 250\,$meV.
This results in a first approximation for the
electron-exchange boson interaction spectral density
$I^2\chi(\omega)$. In a second step
this approximate spectral density is refined using a least squares
fit procedure based on the full non-linear Eliashberg equations \eqref{eq:Bx}.
This second step is of particular importance when inverting superconducting
state data.

\begin{figure}[pt]
  \vspace*{-5mm}
  \includegraphics[width=9cm]{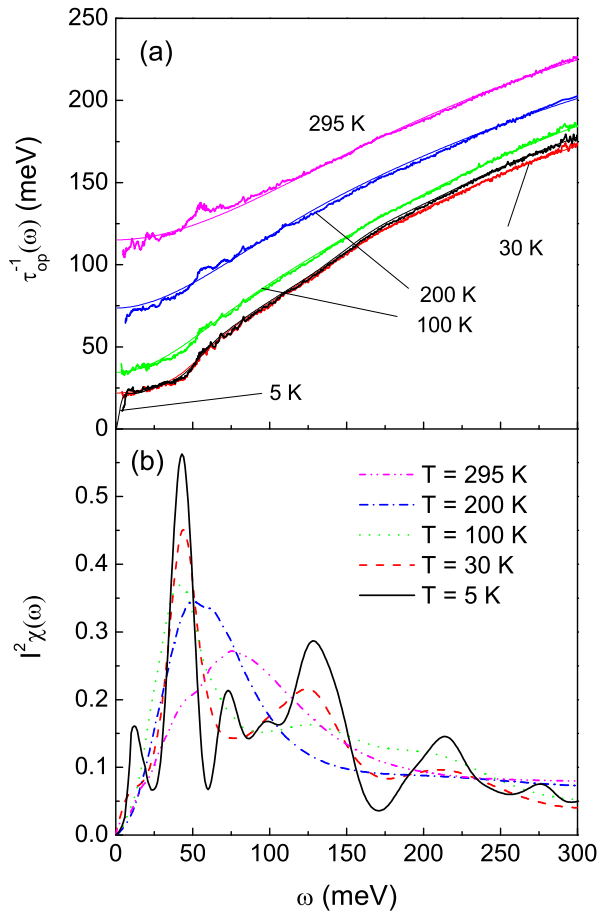}
  \caption{(Color online) (a): The optical scattering rate
$\tau^{-1}_{\rm op}(\omega)$ in meV as a function of $\omega$ (in meV) at five
temperatures, namely $T=300\,$K (magenta, dash-double dotted line),
$200\,$K (blue, dash-dotted line), $100\,$K (green, dotted line), and
$30\,$K (red, dashed line) in the normal
state and $T=5\,$K [black solid line, the only curves with
$\tau^{-1}_{\rm op}(\omega=0) \equiv 0$] in the superconducting state for
the optimally doped PCCO single crystal
($x=0.15$). The heavy lines present experimental data while the light lines
correspond to theoretical results obtained from solutions of the full
Eliashberg equations \eqref{eq:Bx} using the $I^2\chi(\omega)$ spectral
densities shown in the bottom frame of this figure. The impurity parameter was
set to $t^+ = \Gamma^+ = 3.1\,$meV. (b): The electron-exchange boson
interaction spectral density $I^2\chi(\omega)$ obtained from the inversion of
the experimental data presented by the heavy lines in the top frame of this
figure. }
  \label{fig:3}
\end{figure}

Fig.~\ref{fig:3} reports the results of our data analysis.
%
%
Figure~\ref{fig:3}(a) demonstrates the quality of data reproduction when
the $I^2\chi(\omega)$ spectra shown in Fig.~\ref{fig:3}(b) are
used to calculate the optical scattering rate using the Eliashberg equations
\eqref{eq:Bx}. The heavy lines represent the data and the thin lines theory.
For the normal state, $T=295\,$K, $200\,$K, $100\,$K, and $30\,$K Eliashberg
equation \eqref{eq:B2} with $\tilde{\Delta}(\nu+i0^+;\vartheta)\equiv 0$ was
used to calculate the renormalized quasiparticle energies
$\tilde{\omega}(\nu+i0^+)$. In this case the impurity scattering rate
$\tau_{\rm imp}^{-1}$ is only a term which is added to the renormalized
energies. An impurity parameter $t^+ = 3.1\,$meV (see Appendix~\ref{app:A}) was
required for best over all data reproduction. It corresponds to an impurity
scattering rate $\tau^{-1}_{\rm imp} = 2\pi t^+ \approx 19.5\,$meV for all
temperatures. This is in excellent agreement with the zero frequency offset of
$\sim 22\,$meV reported for the $T=30\,$K data.

Particular attention is required for the superconducting state. First
%
%
\begin{figure}[pt]
\includegraphics[width=9cm]{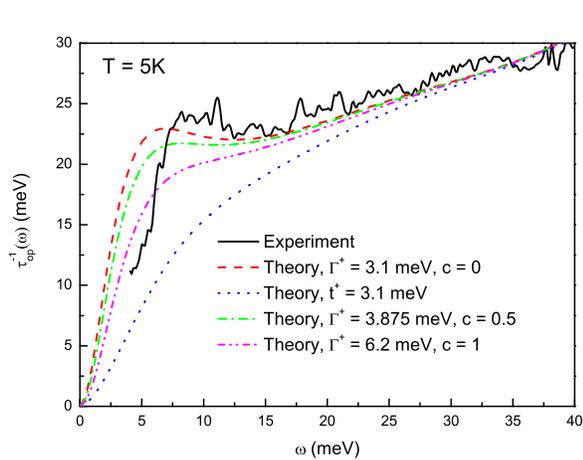}
\caption{(Color online) The solid black line represents the experimental
$\tau^{-1}_{\rm op}(\omega)$ as a function of $\omega$ at $T=5\,$K. The dashed
curve presents the results of an Eliashberg theory calculation using unitary
impurity scattering described by the impurity parameter $\Gamma^+ = 3.15\,$meV.
The dotted line corresponds to Born limit scattering described by the parameter
$t^+=3.15\,$meV. Finally, the dash-dotted curve represents an intermediate case
with $\Gamma^+ = 3.94\,$meV and $c=0.5$ and, furthermore, the dash-double
dotted curve belongs to $\Gamma^+ = 6.3\,$meV and $c=1$. }
\label{fig:4}
\end{figure}
of all, impurities are always pair-breaking in $d$-wave superconductors and
thus effectively reduce the critical temperature of the impure sample in
comparison to the `clean limit' critical temperature. Furthermore, there are
two limits of impurity scattering to be considered. One limit is described by
unitary or resonant scattering which is characterized by the parameter
$\Gamma^+$ in Eq.~\eqref{eq:B2}. The other limit is Born's scattering (or weak
scattering) characterized by the impurity parameter $t^+$. In reality the
scattering law is intermediate between unitary and Born limit scattering. This
is characterized by the parameter $c$ in Eq.~\eqref{eq:B2}; $c=0$ is the
unitary limit and $c\to\infty$ the Born limit. Another complication arises from
the fact that the relevant kernel \eqref{eq:6} does not contain impurity
scattering in contrast to its normal state counterpart Eq.~\eqref{eq:5}.
Furthermore, Eq.~\eqref{eq:6} requires some knowledge of the size of the gap
amplitude $\Delta$. Fortunately, this is not critical in this particular case
because the rather big impurity scattering rate of $\sim 22\,$meV, suggests
that the sample will already be in the gapless regime even at
$T=5\,$K.\cite{ag} Thus, it will be sufficient for the first step, the maximum
entropy deconvolution of Eq.~\eqref{eq:4}, to use that value of $\Delta$ for
which the best data reproduction can be achieved. This procedure results in a
first approximation of the $I^2\chi(\omega)$ spectrum which is then
parameterized and improved by a least squares fit to the data using the
solutions of the full non-linear Eliashberg equations to calculate the optical
scattering rate. In this step the impurity parameter $t^+$ stays unchanged and
only Born scattering is treated but is included in each
iteration of the Eliashberg equations \eqref{eq:Bx} and, of course,
this greatly reduces the value of the critical temperature $T_c$.
The result are the spectral
density $I^2\chi(\omega)$ shown by the solid line in Fig.~\ref{fig:3}(b).

In a final step the low energy regime $0\le\omega\le 10\,$meV in which the
initial slope of $\tau^{-1}_{\rm op}(\omega)$ is dominated by impurity
scattering is to be fitted using full non-linear Eliashberg theory,
Eqs.~\eqref{eq:Bx}, taking into account the different laws of impurity
scattering. As the scattering rate itself has already been determined, the
appropriate scattering law can now be determined by the best possible fit to
the data in this low energy region. Fig.~\ref{fig:4} demonstrates the results
of such a procedure. The solid black line corresponds to the experimental data,
the dashed line represents unitary limit scattering with $\Gamma^+ = 3.1\,$meV
and the dotted line shows the result for Born limit scattering with $t^+ =
3.1\,$meV. For comparison we include additional results for intermediate
impurity scattering, namely $\Gamma^+ = 3.875\,$meV with $c=0.5$ and
$\Gamma^+=6.2\,$meV and $c=1$. Obviously, impurity scattering is much closer to
unitary than to Born type scattering but it is impossible to decide whether a
better fit is found for $c=0$ (unitary scattering) or $c=0.5$ (intermediate
scattering). It also becomes apparent that impurity scattering does not
affect the energy dependence of $\tau^{-1}_{\rm op}(\omega)$ for energies
$\omega > 30\,$meV and, hence, there is no need to get yet a new
estimate for $I^2\chi(\omega)$.
It is certainly interesting to note in passing that the
clean limit (intrinsic) critical temperature $T_{c0}$ for $t^+=0$ is
approximately $61\,$K so that the presence of impurity scattering has
greatly reduced the value of the critical temperature.

We return to Fig.~\ref{eq:3}(b) which presents the spectral
densities $I^2\chi(\omega)$ as a result of the inversion process. Five
temperatures are shown, namely $T=5\,$K (solid line), $30\,$K (dashed line),
$100\,$K (dotted line), $200\,$K (dash-dotted lines), and $295\,$K (dash-double
dotted line). The main feature is a very pronounced peak centered around $\sim
45\,$meV and which gets reduced in amplitude
with increasing temperature. This main
peak is followed by a valley-hump structure which also becomes less pronounced
with increasing temperature an has disappeared at $T=200\,$K.
At $T=295\,$K the $I^2\chi(\omega)$ spectral density takes on the form
of a simple MMP-form as was proposed by Millis {\it et
al.}\cite{millis} for a spin-fluctuation spectrum with its maximum at $\sim
80\,$meV. In these respects the $I^2\chi(\omega)$ spectral density found for the
optimally doped PCCO single crystal follows the pattern found for all other
hole doped High-$T_c$ superconductors analyzed so far. It is, finally,
important to point out that the $T=5\,$K shows an additional peak centered
around $\sim 12.5\,$meV. This peak describes a possible coupling of the charge
carriers to a mode at this energy.

\begin{figure}[pt]
  \includegraphics[width=9cm]{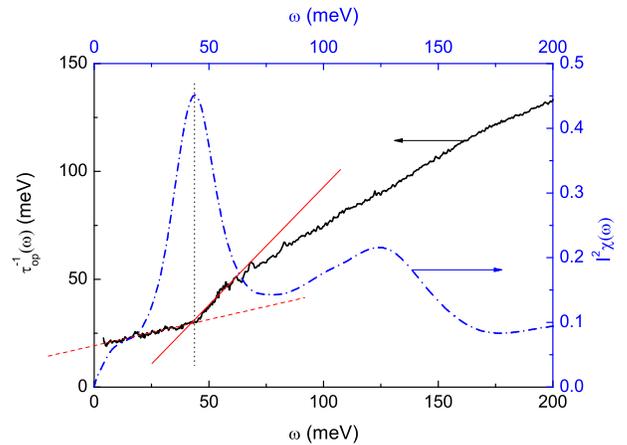}
  \caption{(Color online) The single crystal experimental
optical scattering rate $\tau^{-1}_{\rm op}(\omega)$ as a function of $\omega$
at $T=30\,$K (black solid line). The two straight lines (red, dashed and solid)
are used
to emphasize the change in slope in $\tau^{-1}_{\rm op}(\omega)$ at $\omega =
45\,$meV. The (blue) dash-dotted line corresponds to the spectral density
$I^2\chi(\omega)$ derived from the $\tau^{-1}_{\rm op}(\omega)$ data by
inversion. }
  \label{fig:6a}
\end{figure}

\begin{figure}[pt]
  \vspace*{-5mm}
  \includegraphics[width=9cm]{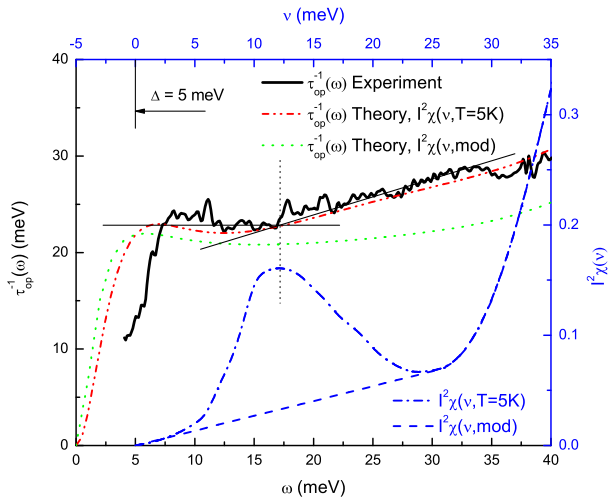}
  \caption{(Color online) The optical scattering rate
$\tau^{-1}_{\rm op}(\omega)$ vs $\omega$ for the optimally doped ($x=0.15$)
PCCO single crystal in the superconducting state at $T=5\,$K. The
(black) heavy solid
line corresponds to experiment. The two (black) light solid lines are
used to emphasize
the change in slope in $\tau^{-1}_{\rm op}(\omega)$ at $\omega\approx 17\,$meV.
The (red) dash-double dotted line represents the theoretical $\tau^{-1}_{\rm
op}(\omega)$ calculated using the spectral density
$I^2\chi(\nu,T=5\,\textrm{K})$ (blue, dash-dotted line) for unitary
impurity scattering with
$\Gamma^+=3.1\,$meV and $c=0$. The (green) dotted line shows the theoretical
$\tau^{-1}_{\rm op}(\omega)$ calculated using the spectral density
$I^2\chi(\nu,\textrm{mod})$ (blue, dashed line) for the same unitary
impurity scattering.
The $\nu$-axis was displaced by the amplitude $\Delta = 5\,$meV of the
superconducting gap at $T=5\,$K with respect to the $\omega$-axis. }
  \label{fig:7}
\end{figure}

We add Figs.~\ref{fig:6a} and \ref{fig:7} to make two important points.
Fig.~\ref{fig:6a} shows the $\tau^{-1}_{\rm op}(\omega)$ data for $T=30\,$K
(solid line) for the single crystal. Superimposed are two straight lines.
%
%
The first, (red) dashed line starts at the $\omega=0$ offset at the value
of the
residual scattering rate $\tau^{-1}_{\rm op}(\omega=0)$ and follows experiment
up to about $45\,$meV. The second, (red) solid line starts near the
origin and
follows the data for energies $> 45\,$meV emphasizing the change in slope at
$\omega=45\,$meV. If the first were perfectly flat (which it is not) the two
straight lines would represent two processes, the Drude (coherent) part of the
scattering rate and the boson assisted (incoherent) part, respectively. If for
the latter we assumed that the boson is an Einstein mode at some energy
$\omega_E$ which is large enough that the boson assisted processes are well
separated from the coherent Drude part, then the coherent part of
$\tau^{-1}_{\rm op} (\omega)$ would be constant and equal to the residual
scattering rate at $\omega=0$ until $\omega=\omega_E$ is reached. At $\omega_E$
the boson assisted absorption sets in as an additional process and
$\tau^{-1}_{\rm op}(\omega)$ develops a kink as is suggested by the intercept
of the two straight lines in Fig.~\ref{fig:6a}. This corresponds precisely to
the peak at the same energy seen in the $I^2\chi(\omega)$ spectrum shown as a
(blue) dash-dotted line in Fig.~\ref{fig:6a} with the horizontal (black)
light dotted line indicating the peak center. The kink in
$\tau^{-1}_{\rm op} (\omega)$ so identified is a clear signature of the peak
at $45\,$meV in $I^2\chi(\omega)$. Obviously, the (red) dashed
straight line has
in our case a non-zero slope as a function of energy. This is the signature of
additional boson assisted processes even at energies $\omega < 45\,$meV and,
consequently, the spectral function $I^2\chi(\omega)$ has non-zero weight at
low energies. Next, we will examine this region more closely.

In Fig.~\ref{fig:7} we address the question whether or not the low
%
%
frequency peak at $12.5\,$meV in the low temperature
$I^2\chi(\nu,T=5\,\textrm{K})$ spectral density (blue, dash-dotted line)
is really
essential for data reconstruction in the region $0\le\omega\le40\,$meV and
reflects features contained in the
$\tau^{-1}_{\rm op}(\omega,T=5\,\textrm{K})$ data.
The (black) heavy solid line represents the data and, as in
Fig.~\ref{fig:6a} we use two (black) thin straight lines to emphasize
the change in slope which takes place at
$\sim 17\,$meV in $\tau^{-1}_{\rm op}(\omega,T=5\,\textrm{K})$.
For comparison we also present
the spectral density $I^2\chi(\nu,\textrm{mod})$ in which the peak
around $12.5\,$meV in spectrum $I^2\chi(\nu,T=5\,\textrm{K})$ has
been replaced by a straight line connecting the origin with the
bottom of the valley at $25\,$meV (blue, dashed line).
As required by theory, the
$\nu$-scale was displaced by the amplitude of the superconducting gap at
$T=5\,$K, $\Delta = 5\,$meV. Following the argument already applied in our
discussion of Fig.~\ref{fig:6a} it becomes obvious that the peak at
$\nu=12.5\,$meV in the spectral density $I^2\chi(\nu,T=5\,\textrm{K})$
coincides with the change of slope of the optical scattering rate at $\sim
17\,$meV as indicated by the intersection of the two (black) thin solid lines.
Thus,
this peak represents, indeed, a real feature of the optical scattering rate and
describes the coupling of the charge carriers to a (weak) mode at $12.5\,$meV.

We added two more curves to this figure. The (red) dash-double dotted curve
represents the
solution of the full non-linear Eliashberg equations \eqref{eq:Bx} found for
the spectral density $I^2\chi(\nu,T=5\,\textrm{K})$ in the superconducting
state at $T=5\,$K. The impurity parameters were set to
$\Gamma^+=3.1\,$meV and $c=0$.
[It is identical to the (black) thin solid line in Fig.~\ref{fig:4}(a).]
We see that this
result follows closely the data over the whole energy region. The second,
(green) dotted curve is also for the superconducting state at $T=5\,$K
but now the
spectral density $I^2\chi(\nu,\textrm{mod})$ has been used. The impurity
parameters remain unchanged. This results falls well below experiment
in the region
$17\le\omega\le 40\,$meV, thus demonstrating the importance of the $12.5\,$meV
peak in $I^2\chi(\nu,T=5\,\textrm{K})$ for best possible data reconstruction.
It is worth noting that the dashed-double dotted and the dashed curves meet
only at higher energies around $70\,$meV. This demonstrates clearly that
a peak in $I^2\chi(\omega)$ centered around the energy $\omega_E$
not only produces a kink in
$\tau^{-1}_\textrm{op}(\omega=\omega_E)$ but is also responsible for increased
scattering at all energies above $\omega_E$.

%
%
\subsection{The \boldmath $x=0.15$ \unboldmath thin PCCO film}

We followed the procedure outlined in the previous subsection to invert the
optical scattering rate $\tau^{-1}_{\rm op}(\omega)$ of an optimally doped,
epitaxially grown thin PCCO film ($x=0.15$, $T_c=21\,$K) described in
Sec.~\ref{sec:2b}. The best over all data reproduction of the normal state
data required an impurity parameter $t^+=5.0\,$meV which corresponds to an
impurity scattering rate $\tau^{-1}_{\rm imp} = 31.4\,$meV ($\sim
253\,$cm$^{-1}$) in excellent agreement with the experimental value of
$32.5\,$meV ($\sim 262\,$cm$^{-1}$) at $T=25\,$K. In the superconducting state
the best agreement was found for an intermediate impurity scattering strength
described by the parameters $\Gamma^+ = 7.8125\,$meV and $c=0.75$.
We also note that in this case the clean limit critical temperature
$T_{c0}\approx 57\,$K. This is a remarkable result as it compares well
with the intrinsic critical temperature of $61\,$K found for the single
crystal. Thus, we conclude that, indeed, both samples present the same
physical properties typical for optimally doped PCCO.

In discussing the results of our data analysis we primarily
%
%
\begin{figure}[pt]
  \vspace*{-5mm}
  \includegraphics[width=9cm]{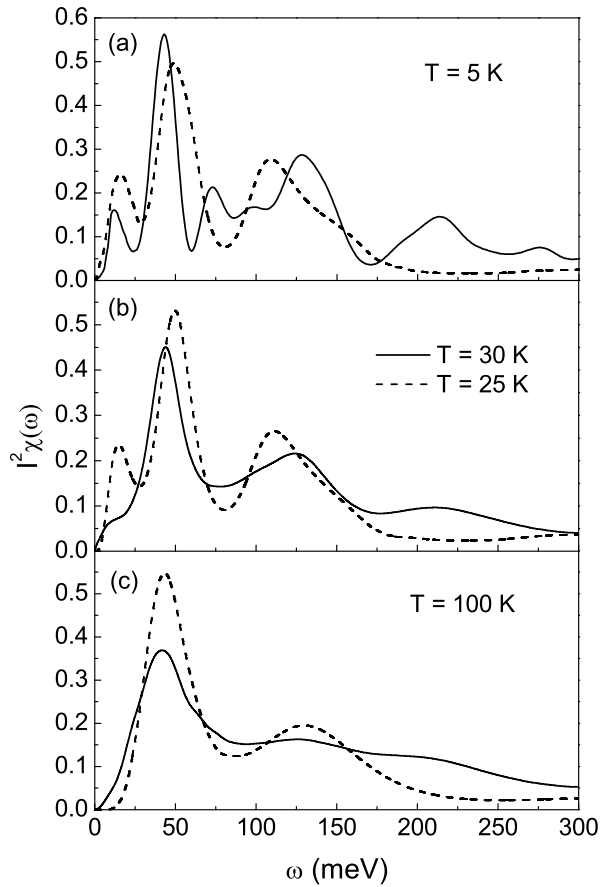}
  \caption{The electron spectral density $I^2\chi(\omega)$ vs $\omega$
for the PCCO $x=0.15$ single crystal (solid lines) and thin film (dashed
lines). The $I^2\chi(\omega)$ of the thin film has been scaled by a factor of
0.5 while the single crystal data is unchanged. This emphasizes similarities
between the two sets of spectra. The results are (a) for the superconducting
state at $T=5\,$K, (b) for the normal state at $T=30\,K$
in the single crystal is compared with the $T=25\,$K spectrum for the thin
film, and (c) for the normal state at $T=100\,$K. }
  \label{fig:6}
\end{figure}
concentrate on the similarities in the electron-boson spectral densities
$I^2\chi(\omega)$ for the two optimally doped samples investigated here. For
this purpose Fig.~\ref{fig:6} presents the $I^2\chi(\omega)$ spectra for the
$x=0.15$ thin film (dashed lines) scaled by a constant factor of 0.5 while the
corresponding spectra of the single crystal (solid lines) stay unchanged.
Figure~\ref{fig:6}(a) is for $T=5\,$K in the superconducting state and
the two spectra reveal very similar features as functions of $\omega$. There is
a first resonance like peak at $\sim 14\,$meV ($\sim12\,$meV in the single
crystal) which corresponds in energy to the spin one resonance reported
by Wilson
{\it et al.}\cite{wilson} and to the peak at the same energy found by STM
experiments\cite{niest} in the system PLCCO. There is a second, main peak at
$50\,$meV in the thin film sample. The single crystal $I^2\chi(\omega)$
spectrum shows this peak at the slightly lower energy of $\sim45\,$meV.
There exists, as yet, no experimental evidence from
neutron scattering for the existence of a corresponding peak in the imaginary
part of the spin susceptibility at such an energy in PCCO and, therefore, our
analysis represents a specific prediction. This second peak is followed by a
valley-hump feature which starts at $\sim80\,$meV with the hump centered at
$\sim107\,$meV. After this structure the spectral density levels off to a
weak background which extends beyond $\omega = 300\,$meV and which has little
structure in the case of the thin film. This result is very similar to what has
been reported recently for the system LSCO by analysis of optical
data\cite{hwang1} (including the valley-hump structure) and by neutron
scattering.\cite{vigno} It is interesting to note an additional valley-peak
structure beginning at $\sim 55\,$meV in the single crystal spectrum
which cannot be observed at higher temperatures. This structure is most likely
due to not completely removed structures in the data around this energy.

Figure~\ref{fig:6}(b) shows the single crystal results for $T=30\,$K (solid line)
and the thin film result for $T=25\,$K (dashed line) in the normal state; both
scaled as described above. While at $T=25\,$K a small signature of the
$14\,$meV peak remains, it has basically vanished at $T=30\,$K in the single
crystal (only a small
shoulder is left). These differences could also be due to the very different
nature of the two samples. The second peak is still very pronounced and
stays basically unchanged in energy in both samples.
The valley-hump structure which follows this second peak at around
$80\,$meV stays unchanged in comparison to the $T=5\,$K spectra as is the
background for $\omega > 200\,$meV.

Finally, Fig.~\ref{fig:6}(c) presents the rescaled spectra for
$T=100\,$K. We see that all structures have been smeared out in the spectra and
the positions of the main peak change only little. The valley-hump structure
which follows the main peak has been smeared out almost completely in the
$I^2\chi(\omega)$ spectral density of the single crystal while it is still well
developed in the thin film spectral density. It is also remarkable that the
thin film spectra have a largely suppressed background for $\omega > 180\,$meV
in comparison to the single crystal spectra. Increasing the temperature further
results in spectra which are MMP-form\cite{millis} like and the maximum at
$T=300\,$K moves to $\sim 80\,$meV in the thin film and to $\sim 90\,$meV in
the single crystal. This corresponds to what has been observed in optimally
doped Bi2212.\cite{hwang}

We also note that the maximum of the hump in the $T=5\,$K and
$25\,$K thin film spectra is at lower energies as compared to the
same structure in the single crystal spectra. At $100\,$K, though,
they agree, i.e.: in the $x=0.15$ thin film spectra the maximum in
the high energy valley-hump structure moves towards higher energies
with increasing temperatures while it stays fixed in the single crystal.
\section{Summary}
\label{sec:4}
Motivated by the report of a spin one resonance in the imaginary part of the
spin susceptibility in the electron-doped system PLCCO at the very low energy
of $\sim11\,$meV by neutron scattering we studied the optical scattering rate
reported for optimally doped PCCO samples using the maximum entropy
technique to extract information on the electron-boson spectral density
$I^2\chi(\omega)$. This spectral density contains information on the coupling
of the charge carriers to bosonic modes. We found that in the superconducting
state the electrons couple to a bosonic mode centered around $12 (14)\,$meV
and that
there is a second higher energy group of modes centered around $45 (50)\,$meV
which has not yet been observed by neutron scattering. Above $T_c$ the
thin film sample still shows some coupling to the low energy mode
at $25\,$K and while in the single crystal no coupling to this mode can
be observed at $T=30\,$K which can certainly be due to the differences
in the two samples. The second,
high energy peak is clearly developed above $T_c$ in all samples and evolves
into an MMP-form like spin fluctuation background with further increasing
temperature. All $I^2\chi(\omega)$ spectra extend to energies $> 300\,$meV.
These results resemble very closely to what has been observed in optimally
doped LSCO samples and, apart from the low energy mode, what has been
reported for optimally Bi2212 samples. All this proves that the electron-doped
system PCCO behaves in its charge dynamics much like all the other hole-doped
High-$T_c$ cuprates. There is one important difference though, PCCO in contrast
to all hole-doped High-$T_c$ superconductors investigated so far, is in the
dirty limit. The residual scattering in all samples investigated here is
sufficiently large to substantially reduce the value of the critical
temperature over its pure (intrinsic) limit. This fact is in full agreement
with results reported by Dagan {\it et al.}\cite{dagan} from their analysis of
PCCO-lead tunneling junctions.

\section*{Acknowledgment}
Research supported in part by the Natural Sciences and Engineering Research
Council of Canada (NSERC) and by the Canadian Institute for Advanced Research
(CIFAR). We thank N. Bontemps for valuable discussions and her keen interest.
JPC and ES want to thank T. Timusk for his interest and
many discussions. ES enjoyed the hospitality and friendship of the members of
the Department of Physics and Astronomy during his visit at McMaster
University. The Work at Brookhaven National Laboratory was supported by the
Office of Science, U.S. Department of Energy, under Contract No.
DE-AC02-98CH10886.

%
%
%
\appendix*
\section{$d$-wave Eliashberg equations for impure systems}
\label{app:A}
The generalization to a $d$-wave gap has already been published by Jiang {\it
et al.}\cite{jiang} and has been used to describe various aspects of the
superconducting state in the cuprates. In the mixed representation of Marsiglio
{\it et al.}\cite{mars2} they are of the form
\begin{widetext}
\begin{subequations}
\label{eq:Bx}
\begin{eqnarray}
  \label{eq:B1}
    \tilde{\Delta}(\nu+i0^+;\vartheta) &=& \pi Tg
  \sum\limits_{m=0}^\infty\cos(2\vartheta)\left[\lambda(\nu-i\omega_m)+
  \lambda(\nu+i\omega_m)\right]h_+(i\omega_m)\\
 &&+i\pi g\int\limits^\infty_{-\infty}\!dz\,\cos(2\vartheta)
  I^2\chi(z)\left[n(z)+f(z-\nu)\right]h_-(i\omega_m\to\nu-z+i0^+),\nonumber
\end{eqnarray}
and, in the renormalization channel,
\begin{eqnarray}
  \tilde{\omega}(\nu+i0^+) &=& \nu+i\pi T%
  \sum\limits_{m=0}^\infty\left[\lambda(\nu-i\omega_m)-
  \lambda(\nu+i\omega_m)\right]g_+(i\omega_m)\nonumber\\
  &&+i\pi\int\limits^\infty_{-\infty}\!dz\,
   I^2\chi(z)\left[n(z)+f(z-\nu)\right]g_-(i\omega_m\to\nu-z+i0^+)\nonumber\\
  &&+i\pi\Gamma^+\frac{g_-(i\omega_n\to\nu+i0^+)}
  {c^2+g^2_-(i\omega_n\to\nu+i0^+)+h^2_-(i\omega_n\to\nu+i0^+)}.
  \label{eq:B2}
\end{eqnarray}
\end{subequations}
Here
\[
 h_\pm(i\omega_m) = \left\langle
 \frac{\tilde{\Delta}(i\omega_m;\vartheta)\cos(2\vartheta)}
  {\sqrt{\tilde{\omega}^2(i\omega_m)\pm\tilde{\Delta}^2(i\omega_m;
  \vartheta)}}\right\rangle_\vartheta,\quad
 g_\pm(i\omega_m) = \left\langle
    \frac{\tilde{\omega}(i\omega_m)}
  {\sqrt{\tilde{\omega}^2(i\omega_m)\pm\tilde{\Delta}^2(i\omega_m;
  \vartheta)}}\right\rangle_\vartheta,
\]
\end{widetext}
and the parameter $g$ in Eq.~\eqref{eq:B1} allows for a possible
difference in spectral
density between the $\tilde{\omega}$ and $\tilde{\Delta}$ channels.
It is fixed to get the measured value of the critical temperature.
In the above $\tilde{\Delta}(i\omega_m;\vartheta)$ is the pairing
energy which its evaluated at the fermionic Matsubara frequencies
$\omega_m = \pi T(2m-1), m = 0,\,\pm1,\,\pm2,\ldots$ and
$\tilde{\omega}(i\omega_m)$ are the renormalized frequencies
evaluated at the same Matsubara frequencies; $f(z)$
and $n(z)$ are the Fermi and Bose distribution, respectively.
Furthermore, the $\vartheta$ dependence of the pairing energy is
described by $\tilde{\Delta}(i\omega_m,\vartheta) =
\tilde{\Delta}(i\omega_m)\cos(2\vartheta)$ with $\vartheta$ the polar
angle in the two-dimensional CuO Brillouin zone. The
brackets $\langle\cdots\rangle_\vartheta$ are the angular average
over $\vartheta$, and
  $\lambda(\nu) = \int^\infty_{-\infty}\!d\Omega\,
   \alpha^2F(\Omega)/(\nu-\Omega+i0^+)$.
Eqs.~\eqref{eq:Bx} are a set of nonlinear coupled equations for the
renormalized pairing potential $\tilde{\Delta}(\nu+i0^+;\vartheta)$
and the normalized frequencies $\tilde{\omega}(\nu+i0^+)$ with
the gap
  $\Delta(\nu+i0^+;\vartheta) = \tilde{\Delta}(\nu+i0^+;\vartheta)/
  Z(\nu)$,
where the renormalization function $Z(\nu)$ was introduced in the
usual way as $\tilde{\omega}(\nu+i0^+) = \nu Z(\nu)$. Finally,
$\tilde{\Delta}(i\omega_n,\vartheta)$ and
$\tilde{\omega}(i\omega_n)$ are the solutions of the equivalent
equations formulated on the imaginary axis.\cite{schachx}

Impurity scattering is described by the term proportional to $\Gamma^+$ in
Eq.~\eqref{eq:B2} and enters only this equation because we assume a pure
$d$-wave model for the pairing potential with zero average over the Fermi
surface while the impurity scattering is assumed to be isotropic. Here,
$\Gamma^+$ is proportional to the impurity concentration and $c$ is related to
the electron phase shift for scattering off the impurity. For unitary or
resonant scattering $c$ is equal to zero while $c\to\infty$ gives the Born
approximation, i.e.: the weak scattering limit. In this limit the entire
impurity term reduces to the form $i\pi t^+ g(i\omega_n\to\nu+i0^+)$ with $c$
absorbed into the impurity parameter $t^+$. In the normal state
$\tilde{\Delta}(\nu+i0^+;\vartheta)\equiv 0$ and there is no need to
distinguish any longer between unitary and Born limit impurity scattering. The
scattering term reduces to $i\pi
t^+\text{sgn}[\tilde{\omega}(i\omega_n\to\nu+i0^+)]$. At the critical
temperature linearized Eliashberg equations are valid, i.e.:
$h(i\omega_n)\simeq 0$ and $g(i\omega_n)\simeq
\text{sgn}[\tilde{\omega}(i\omega_n)]$. Thus, at $T_c$,
$t^+\{c^2+\text{sgn}[\tilde{\omega}(i\omega_n)]\} = \Gamma^+$ and this relates
immediately $\Gamma^+$ to the impurity scattering rate via $\tau^{-1}_{\rm imp}
= 2\pi t^+$.

The optical conductivity follows from knowledge of $\tilde{\omega}$
and $\tilde{\Delta}$. The formula to be evaluated is
\begin{widetext}
\begin{equation}
  \sigma_{op}(T,\nu) = \frac{\Omega^2_p}{4\pi}\frac{i}{\nu}
     \left\langle
      \int\limits_0^\infty\!d\omega\,\tanh\left(\frac{\beta\omega}{2}
        \right)\left[
        J(\omega,\nu)- J(-\omega,\nu)
      \right]\right\rangle_\vartheta.  \label{eq:B4}
\end{equation}
The function $J(\omega,\nu)$ is given by
%
\begin{equation}
  \label{eq:B6}
  2J(\omega,\nu) = \frac{1-N(\omega;\vartheta)N(\omega+\nu;\vartheta)-
   P(\omega;\vartheta)P(\omega+\nu;\vartheta)}{E(\omega;\vartheta)+
   E(\omega+\nu;\vartheta)}
   +\frac{1+N^\ast(\omega;\vartheta)N(\omega+\nu;\vartheta)+
   P^\ast(\omega;\vartheta)P(\omega+\nu;\vartheta)}{E^\ast(\omega;\vartheta)-
   E(\omega+\nu;\vartheta)},
\end{equation}
\end{widetext}
with
  $E(\omega;\vartheta) = \sqrt{\tilde{\omega}^2
   (\omega+i0^+)-\tilde{\Delta}^2
   (\omega+i0^+;\vartheta)}$,
  $N(\omega;\vartheta) = \tilde{\omega}(\omega+i0^+)/
   E(\omega;\vartheta)$, and
  $P(\omega;\vartheta) = \tilde{\Delta}(\omega+i0^+;\vartheta)/
   E(\omega;\vartheta)$.
Finally, the star refers to the complex conjugate.

%
%
\bibliography{pccoz}
\end{document}